\documentclass[useAMS,usenatbib]{mn2e}

\usepackage{graphicx}

\title[The v$_c$-$\sigma_c$ relation in low mass and low surface
brightness galaxies]{The v$_c$-$\sigma_c$ relation in low mass and low
surface brightness galaxies} \author[P. Buyle et
al.]{P. Buyle$^{1,2}$\thanks{E-mail:
Pieter.Buyle@UGent.be}\thanks{Post-doctoral Fellow of the Fund for
Scientific Research - Flanders, Belgium (F.W.O.)}, L. Ferrarese$^{2}$,
G. Gentile$^{3}$, H. Dejonghe$^{1}$, M. Baes$^{1}$, U. Klein$^{4}$\\
$^{1}$Sterrenkundig Observatorium, Universiteit Gent, Krijgslaan 281,
S9, B-9000, Ghent, Belgium\\ $^{2}$Herzberg Institute of Astrophysics,
National Research Council of Canada, Victoria, BC V9E2E7, Canada\\
$^{3}$University of New Mexico, Department of Physics and Astronomy,
800 Yale Blvd NE, Albuquerque, New Mexico 87131, USA\\ $^{4}$Bonn
University, Auf dem H\"ugel 71D, 53121 Bonn, Germany }

\def\farcs{\hbox{$.\!\!^{\prime\prime}$}}
\def\kms{km s$^{-1}$}
\def\vs{$v_c - \sigma_c$}
\def\mnras{MNRAS}
\def\aj{AJ}
\def\apj{ApJ}
\def\aap{A\&A}
\def\apjs{ApJS}
\def\apjl{ApJ}

\begin{document}

\date{Accepted 1988 December 15. Received 1988 December 14; in original form 1988 October 11}

\pagerange{\pageref{firstpage}--\pageref{lastpage}} \pubyear{2002}

\maketitle

\label{firstpage}

\begin{abstract}
We present an updated investigation of the relation between large
scale disk circular velocity, $v_c$, and bulge velocity dispersion,
$\sigma_c$. New bulge velocity dispersions are measured for a sample
of 11 low surface brightness (LSB) and 7 high surface brightness (HSB)
spiral galaxies for which $v_c$ is known from published optical or HI
rotation curves.  We find that, while LSB galaxies appear to define
the upper envelope of the region occupied by HSB galaxies (having
relatively larger $v_c$ for any given $\sigma_c$), the distinction
between LSB and HSB galaxies in the \vs~plane becomes less pronounced
for $\sigma_c \la 80$ km s$^{-1}$. We conclude that either the scatter
of the \vs~ relation is a function of $v_c$ (and hence galaxy mass) or
that the character of the \vs~relation changes at $v_c \sim 80$
\kms. Some inplications of our findings are discussed.
\end{abstract}


\section{Introduction}
The existence of supermassive black holes (SBHs) in galactic nuclei
has been accepted since the detections in the nearby spiral NGC 4258
\citep{miyoshi} and in our own galaxy \citep{schodel,ghez}. Compelling
cases now exist in three dozen additional galaxies (see Ferrarese \&
Ford 2005 for a review), making it possible to investigate the
existence of scaling relations linking the SBH mass, $M_{BH}$, to the
overall properties of the host.  Indeed, $M_{BH}$ correlates with the
blue luminosity $L_B$ of the host bulge \citep{kormendy,marconi}, with
the central light concentration \citep{graham}, and with the bulge
velocity dispersion $\sigma_c$ \citep{laura2,geb}.  Because of its
tight scatter (0.34 dex in $M_{BH}$), the $M_{BH}-\sigma_c$ relation
is thought to allow us to peer into the mechanisms controlling the
joint formation/evolution of SBHs and galaxies.  However, cosmological
simulations show that a possibly more fundamental relation should be
expected between $M_{BH}$ and the total gravitational mass $M_{tot}$
of the galaxy, or the mass $M_{DM}$ of the dark matter (DM) halo
\citep[e.g.][]{tiziana,kawakatu05,wyithe05,adams,monaco,haehnelt1,cattaneo,silk,haehnelt2,loeb}.
Indirect observational evidence for such a link has been found in the
form of a correlation between the circular velocity $v_c$ of spirals'
disks (obtained by either HI or deep optical observations) and the
$\sigma_c$ of their bulge component (Ferrarese 2002; Baes et al. 2003;
Pizzella et al. 2005):

\begin{equation}
\log{v_c}=(0.84\, \pm\, 0.09)\log{\sigma_c} + (0.55\, \pm\, 0.29)
\end{equation}

Further exploring the $v_c-\sigma_c$ relation is of interest because,
by linking galactic components on vastly different scales, the
relation is a reflection of the interplay between disks and spheroids
during galaxy formation/evolution.  Furthermore, if $v_c$ and
$\sigma_c$ are used as surrogates for $M_{DM}$ and $M_{BH}$
respectively, equation (1) implies a strong causal relation between
SBHs and DM haloes.

Pizzella et al. (2005) report evidence of a separation in the
$v_c-\sigma_c$ plane between High and Low Surface Brightness spiral
galaxies (HSB and LSB respectively), the latter having distinctly
larger $v_c$ at a given $\sigma_c$ compared to the former. Based on
this, the authors argue against the importance of baryonic collapse in
shaping the density profiles of DM haloes in LSBs, and suggest that
for a given DM halo mass, LSBs might host less massive SBHs than HSBs.
The behaviour of HSBs themselves is somewhat controversial: while
Pizzella et al. (2005) find that HSB in the range $50 \la \sigma_c \la
350$ km\, s$^{-1}$ (corresponding to $150 \la v_c \la 550$ km\,
s$^{-1}$) follow a linear relation, Ferrarese (2002) and Baes et
al. (2003) claim that the relation is mildly non-linear for $\sigma_c
\ga 80$\, km\, s$^{-1}$ ($v_c \ga 140$\, km\, s$^{-1}$),  and breaks
down altogether at lower $\sigma_c$, possibly revealing the inability
of the least massive haloes to form central SBHs
\citep{shankar06,haehnelt1,silk,loeb}

The goal of this paper is to investigate the $v_c-\sigma_c$ relation
for a combination of LSB and HSB galaxies, with $v_c$ in the critical range
$110-220$ km\, s$^{-1}$. The data reduction and analysis are discussed
in \S2 and \S3 respectively. Discussion and conclusions can be found
in \S4.

\begin{figure}
\includegraphics[angle=0,width=8.5cm]{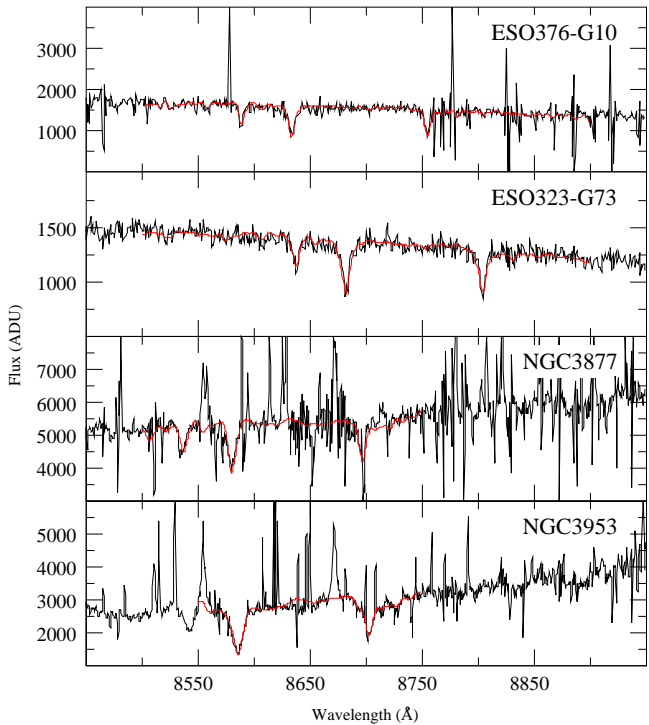}
\caption{Spectra for four of the program galaxies are shown in black.
The top two panels show spectra obtained at the VLT, while the
bottom two panels show Calar Alto spectra. To illustrate the quality of the
data, for each telescope we chose to show the two spectra with the
lowest (first and third panel from the top) and highest (second and
fourth panel from the top) signal-to-noise ratio. The red lines show
the best $\chi^2$ fits to each spectrum, overplotted to the real data in
the spectral region used to perform the fits. The narrow emission
features present in three of the spectra are artifacts resulting from
imperfect sky subtraction}
\end{figure}
\label{ppxf}
\section{Observations and Data Reduction}

To compare the behaviour of LSBs and HSBs in the
$v_c\la220$\, km s$^{-1}$ range of the $v_c-\sigma_c$ relation,
we drafted a sample of galaxies with $v_c$ known from published
rotation curves, and proceeded to obtain new optical spectra from
which to measure $\sigma_c$. We targeted 11 LSB galaxies with optical
rotation curves from the Palunas \& Williams (2000) compilation, and
seven HSB galaxies with HI rotation curves from the Ursa Major sample
of Verheijen (2001).  Besides the need of selecting galaxies visible
in the given observing season, and bright enough to produce spectra
with the required signal-to-noise ratio (S/N), the one essential
criterium in our sample selection is that the published rotation
curves must be symmetric relative to the galaxy's centre, and reach an
asymptotic value at large radii.

Details of the sample and observations are given in Table 1.  The LSB
galaxies were observed with the VLT/UT4 telescope of the European
Southern Observatory on the nights of 2004 April 20, 29 and May 6. The
holographic grism GRIS\_1028z+29 on FORS2 was centred on the Calcium
absorption triplet around 8500\AA, producing an instrumental
broadening $\sigma_{instr}\approx 35$\, km\, s$^{-1}$ with the
0\farcs7 slit. The seven HSB galaxies were observed with the TWIN
spectrograph at the 3.5m Calar Alto telescope, on 2005, June 6-19.
The T06 grating was used in 1st order, again centred at the Ca
triplet. The instrumental broadening was $\sigma_{instr}\approx20$\,
km\, s$^{-1}$ for the 1\farcs2 wide slit. All spectra were divided
into several exposures to ease cosmic ray identification and removal.

Standard data reduction was performed with MIDAS, IRAF and additional
software developed specifically for this project.  After basic data
processing (dark and bias subtraction, flat fielding  and trimming),
atmospheric emission lines were identified and removed through an
interpolation scheme, after which the spectra were rectified and
wavelength calibrated.  Cosmic rays were removed by applying a median
filter along the spatial axis.

\begin{figure}
\includegraphics[angle=0,width=9cm]{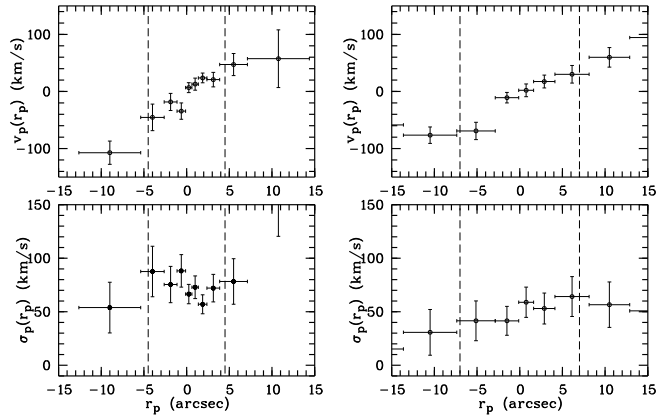}
\caption{Velocity and velocity dispersion curves for ESO215-G39 (left
panel) and ESO376-G10 (right panel) derived by the direct $\chi2$
method. The vertical dashed lines are drawn at the location of the
bulge effective radius, $r_e$.}
\label{profile}
\end{figure}

\begin{figure*}
\begin{center}
\includegraphics[angle=0,width=8.5cm]{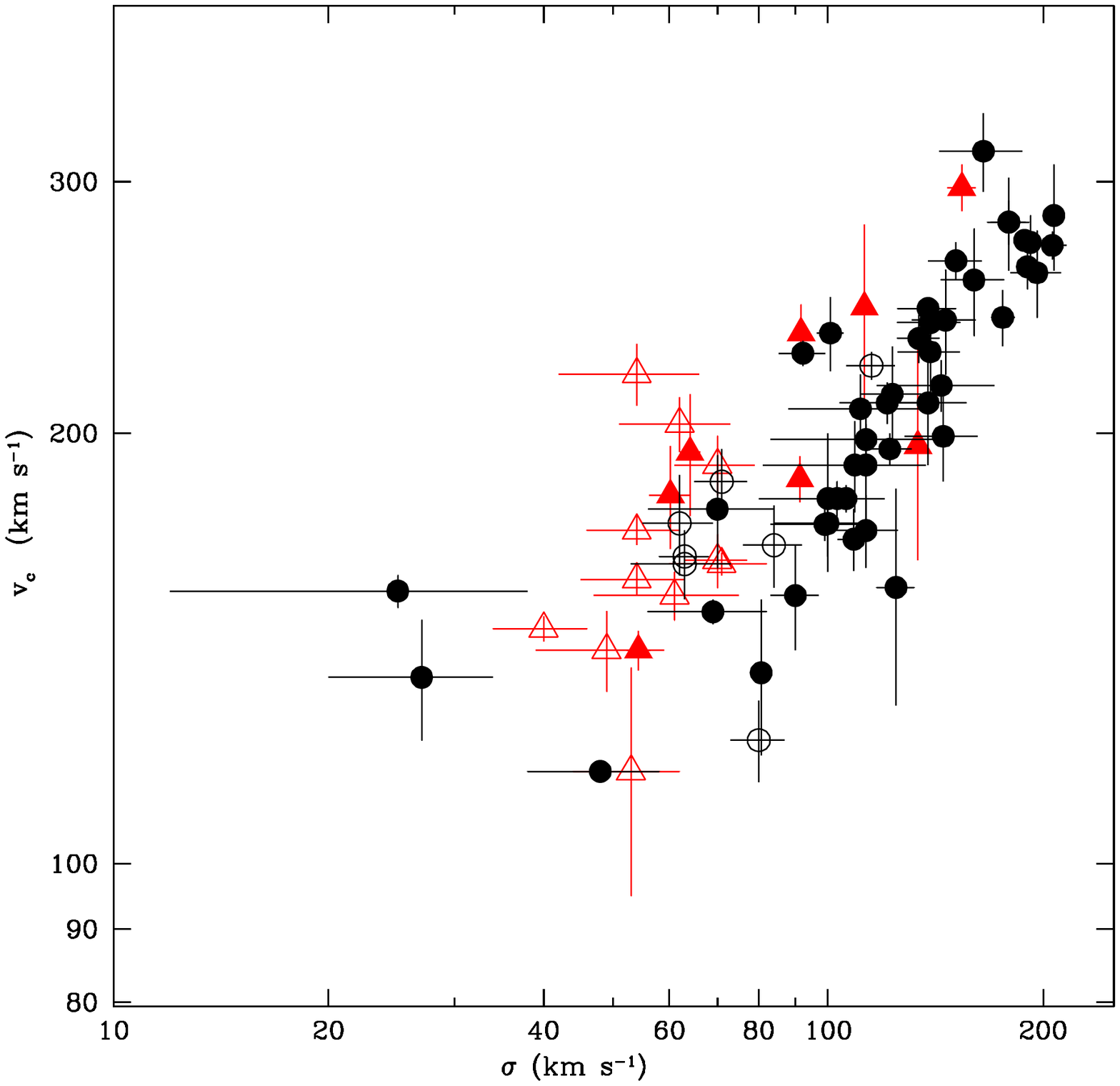}\hspace*{0.5cm}
\includegraphics[angle=0,width=8.5cm]{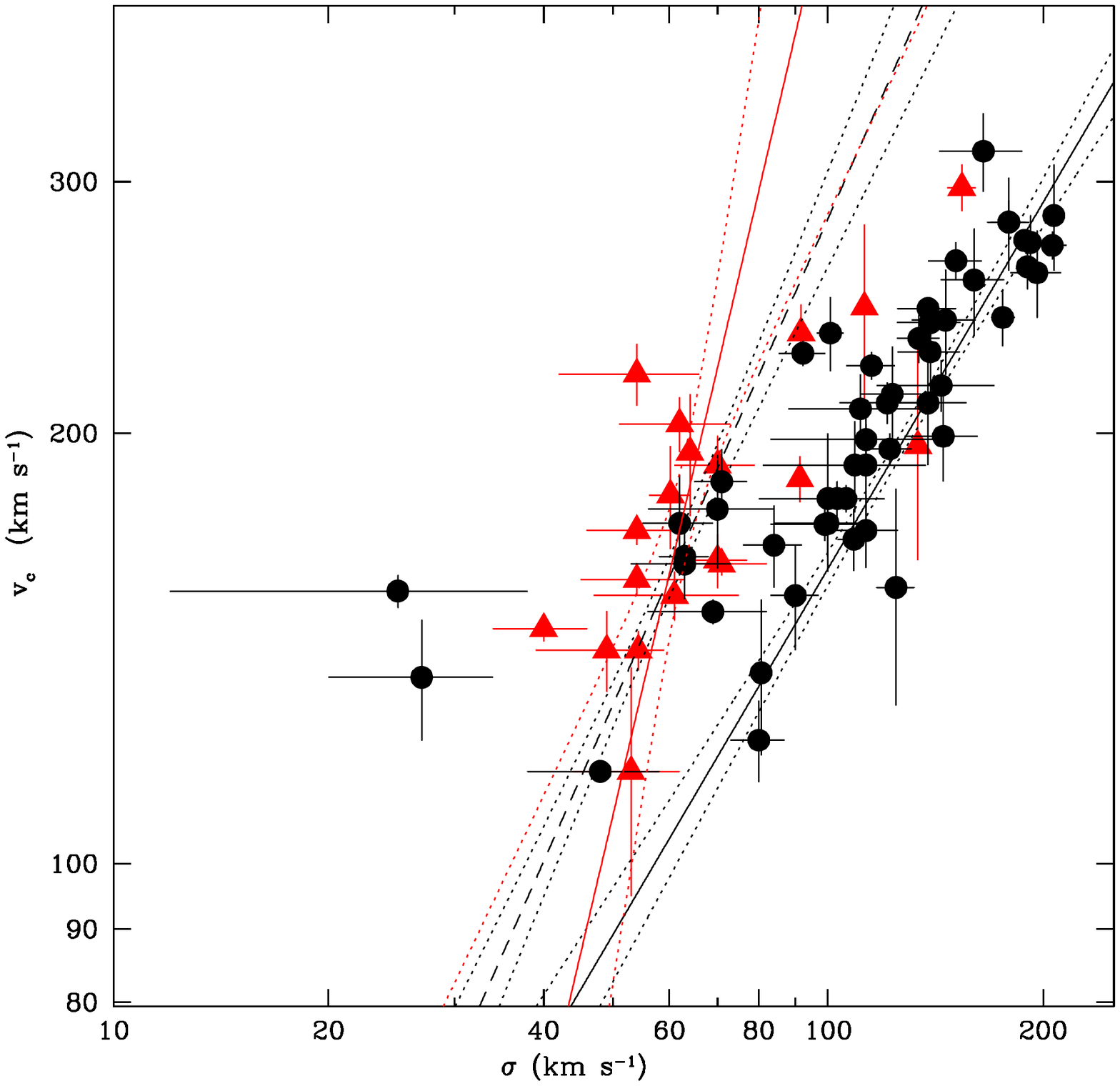}
\caption{Left panel: The relation between the large scale disk
circular velocity $v_c$ and the central bulge velocity dispersion
$\sigma_c$ for HSBs (black circles) and LSBs (red triangles). Data from
Ferrarese (2002), Baes et al. (2003), Pizzella et al. (2005) (full
symbols) and this paper (open symbols) are plotted.  Right panel: as
in the left panel, with fits superimposed. The solid and dashed black
lines are least square fits to HSB galaxies with $\sigma_c \geq 80$
\kms~ and $\sigma_c \leq 80$ \kms~ respectively. The red solid line is
the fit to LSB galaxies with $\sigma_c \leq 80$ \kms.  Dotted lines
represent the 1$\sigma$ uncertainties in the fits.\label{vs}}
\end{center}
\end{figure*}

\section{Data analysis}

Deriving accurate kinematical quantities of galaxies is a non-trivial
task. The result of the extraction can vary depending on the applied
fitting techniques, the S/N, the template stars, the absorption lines
used, etc. (e.g., De Bruyne et al. 2003). To lower this degree of
subjectivity, we derived velocity dispersions using two independent
techniques: a direct $\chi^2$ fit to the spectrum \citep{veronique2},
and the PPXF algorithm \citep{cappellari} (see Fig.~\ref{ppxf}), which
is based on a Penalized Likelihood approach. 
Velocity dispersions derived using these two methods agree to
within 10\%;  the final values quoted in Table 1 were
obtained using the  $\chi^2$ routine.  In analyzing the spectra,
sky residuals that deviated by
more than 3 times the average sigma in the nearby continuum were
rejected; furthermore, to minimize the effects of template mismatch,
six different template stars were used, each with weight calculated by
a quadratic fit to the central row of the spectrum. In keeping with
previous works, velocity dispersions were extracted from spectra
binned, in the spatial direction, out to 1/8th of the bulge effective
radius, estimated from the literature in the case of the HSB galaxies
\citep{baggett,courteau,heraudeau,mollenhoff},  and from published
surface brightness profiles for the Palunas \& Williams (2000)
galaxies.  Within this radius,  the effect of a possible contamination
by the disk rotational velocity should be negligible. This was
verified by measuring the dispersion profiles along the major axis
(without binning spatially); the velocity dispersion of most galaxies
(see Fig.~\ref{profile}) indeed remained nearly constant in the bulge
dominated region.

The circular rotational velocities and errors are published by
Verheijen (2001) for the HSB galaxies. These were derived from HI
observations and extend beyond the optical radius ($R_{25}$), well into
the flat part of the curve. For the LSB galaxies, we measured $v_c$ as
the weighted average of the flat part of the optical rotation curve
published in Palunas \& Williams (2000); the error on $v_c$ was
derived by means of a bootstrap method and reflects any slight
asymmetries in the rotation curves. The optical rotation curves are
not as extended as the HI observations of the HSB galaxies, however
all rotation curves reach an asymptotic value at large radii.  As
shown in Pizzella et al. (2005), when this condition is satisfied,
$v_c$ can be  extracted reliably; therefore we do not expect any
systematic biases in the $v_c$ measurements between our samples of
HSBs and LSBs.  The results of the dynamical analysis are given in
Table 1.

\section{Discussion and Conclusions}
The $v_c - \sigma_c$ relation for HSB galaxies is plotted in the left
panel of Fig.~\ref{vs}. The 11 LSB galaxies studied in this paper
populate the region between $40 \leq \sigma_c \leq 80$ \kms, where
previously only three galaxies could be found. Furthermore, in the
same region, we more than double the number of HSB galaxies, from
three to eight. Pizzella et al. (2005) conclude that, relative to the
\vs~relation defined by HSB galaxies, LSB galaxies appear to have
larger circular velocities, $v_c$, at any given value of
$\sigma_c$. This conclusion is reaffermed by our data.  However, with
the addition of the present sample, it has now become more evident
that there is some overlap in the locations occupied by LSB and HSB
galaxies in the \vs~plane. In other words, LSB galaxies appear to
occupy the {\it upper envelope} of the strip defined by HSB
galaxies. Furthermore, at low velocity dispersions $30 \leq \sigma_c
\leq 80$ \kms, the \vs~relation for both LSB and HSB galaxies appears
to be shifted towards larger circular velocities relative to the
relation defined by the (HSB) galaxies with larger velocity
dispersion.

A least-square fit, taking into account the errors on both quantities
\citep[FITEXY;][]{press92}, to the sample of HSBs with $\sigma_c \geq 80$
\kms~gives:

\begin{equation}
\log{v_c}=(0.68 \pm 0.04)\log{\left(\frac{\sigma_c}{129 {\rm~km~s^{-1}}}\right)+(2.33 \pm 0.01)},
\end{equation}

\noindent with $\chi^2_r=2.58$. For $30 \leq \sigma_c \leq 80$ \kms, on the other hand,

\begin{equation}
\log{v_c}=(1.33 \pm 0.86)\log{\left(\frac{\sigma_c}{63 {\rm~km~s^{-1}}}\right)+(2.21 \pm 0.03)},
\end{equation}

\noindent with $\chi^2_r=0.23$ for HSB galaxies, and 

\begin{equation}
\log{v_c}=(1.91 \pm 0.68)\log{\left(\frac{\sigma_c}{58 {\rm~km~s^{-1}}}\right)+(2.20 \pm 0.02)},
\end{equation}

\noindent with $\chi^2_r=1.55$ for LSB galaxies.

Although the uncertainties are large, equations (3) and (4) are
consistent with each other well within 1$\sigma$, but inconsistent
with equation (2) (see also right panel of Figure~\ref{vs}). 

Two interpretations are open at this point. The samples studied to
date are far from complete or homogeneous, and it is conceivable that,
as more data are collected, more galaxies will fill the gap between
the LSB galaxies, which are mainly found in the low-velocity
dispersion region, and the extrapolation to low $\sigma_c$ of equation
(2), defined mainly by HSB galaxies. This interpretation allows for
the possibility that the log-log \vs~relation, including both HSB and
LSB galaxies, is linear, but that its {\it scatter} increases going
from large to small velocity dispersions. Alternatively, it is
possible that the log-log \vs~relation is not linear, and that indeed
its character changes dramatically below $v_c \sim 200$ km s$^{-1}$,
as originally suggested by Ferrarese (2002) and Baes et al. (2003). In
either case, it seems unavoidable to conclude that the efficiency of
bulge (and, perhaps, SBH) formation, is regulated by intrinsic
parameters in addition to the depth of the global potential well.

The implications of these findings for the central SBHs cannot be
easily quantified. The $M_{BH} - \sigma_c$ relation is not
characterised for LSBs, for which no SBH detection has been
attempted. It is also not defined below $\sigma_c \leq 80$ km
s$^{-1}$, which corresponds to SBHs too small to be detected given
current space or ground-based instrumentation. We reiterate the
conclusion of Ferrarese et al. (2002) and Baes et al. (2003) that the
behaviour of the $v_c - \sigma_c$ relation at $\sigma_c \la 80$ km
s$^{-1}$ might reflect the inability of these galaxies to form a
central SBH, as argued on theoretical grounds by, for instance,
\cite{shankar06,haehnelt1,silk,loeb}.  Numerical simulations
\citep{tiziana,robertson} predict that the $M_{BH}-\sigma_c$ relation
originates from the feedback between the central SBH and the
progenitor of the hot stellar component at the early stages of the
formation of both; therefore, in low-mass, late-type HSB and LSB
galaxies internal factors might lead to the suppression of such
feedback, creating a dynamical relation between disk, bulge and SBH
which is starkly different from that displayed by more massive
systems.

\begin{table*}
{\scriptsize
\begin{tabular}{lcclrrrlrcrc}
\hline
Name & R.A. & Dec. &  $v_c$ & $\sigma_c$ & $r_e$ & $v_{rad}$ &  Morph & $m_I$ & Incl. & Exp.Time & S/N \\
& (J2000) & (J2000) & (km s$^{-1}$) & (km s$^{-1}$) & (arcsec) & (km s$^{-1}$) & & (mag) & (degrees) & (seconds) & (@ 8500 \AA)\\
\hline
\multicolumn{12}{c}{}\\
\multicolumn{12}{c}{LSB galaxies}\\
\multicolumn{12}{c}{}\\
\hline
ESO215-G39 & 11h17m04s & $-$49$^{\circ}$12$'$05$''$ & 162$\pm$3 & 71$\pm$11\phantom{0} & 4.5 & 4335 & SABc & 12.01 & 50 & 500  & 20.03\\
ESO268-G44 & 12h48m42s & $-$45$^{\circ}$00$'$29$''$ & 190$\pm$9 & 70$\pm$9\phantom{0} &  3.5 & 3477 & Sb & 12.22 & 62 & 450  & 25.68\\
ESO322-G19 & 12h29m07s & $-$40$^{\circ}$40$'$24$''$ & 141$\pm$9 & 49$\pm$10\phantom{0} & 0 & 3100 & SBc & 12.64 & 79 & 900  & 22.19\\
ESO323-G42 & 12h55m01s & $-$40$^{\circ}$58$'$20$''$ & 158$\pm$4 & 54$\pm$9\phantom{0} & 4 & 4203 & Sc & 11.53 & 69 & 450  & 30.32\\
ESO323-G73 & 13h04m02s & $-$38$^{\circ}$11$'$57$''$ & 163$\pm$7 & 70$\pm$7\phantom{0} & 2 & 4929 & Sbc & 12.44 & 48 & 450  & 35.29\\
ESO374-G03 & 09h51m57s & $-$33$^{\circ}$04$'$25$''$ & 146$\pm$3 & 40$\pm$6\phantom{0} & 6 & 2931 & SABc & 11.51 & 71 & 600  & 21.80\\
ESO382-G06 & 13h05m32s & $-$32$^{\circ}$57$'$38$''$ & 171$\pm$4 & 54$\pm$8\phantom{0} & 2 & 4809 & Sab & 13.20 & 54 & 600  & 21.01\\
ESO444-G21 & 13h23m31s & $-$30$^{\circ}$06$'$52$''$ & 116$\pm$21 & 53$\pm$9\phantom{0} & 4.5 & 4265 & Sc & 12.86 & 84 & 1900  & 20.46\\
ESO444-G47 & 13h28m25s & $-$31$^{\circ}$51$'$28$''$ & 154$\pm$6 & 61$\pm$14\phantom{0} & 0 & 4389 & SBc & 12.82 & 71 & 900  & 22.01\\
ESO509-G91 & 13h40m03s & $-$25$^{\circ}$28$'$28$''$ & 203$\pm$9 & 62$\pm$11\phantom{0} & 6.5 & 5113 & SBc & 12.70 & 79 & 900  & 20.91\\
ESO376-G10 &10h42m00s & $-$36$^{\circ}$56$'$07$''$ & 220$\pm$11 & 54$\pm$12\phantom{0} & 7 & 3184 & SBd & 11.00 & 76 & 600  & 20.27\\
\hline
\multicolumn{12}{c}{}\\
\multicolumn{12}{c}{HSB galaxies}\\
\multicolumn{12}{c}{}\\
\hline
NGC 3953 & 11h53m49s& +52$^{\circ}$19$'$36$''$ & 223$\pm$5& 115$\pm$9 & 66 & 1052 & Sbc & 9.03 & 63 & 2500  & 60.01 \\
NGC 3877 & 11h46m08s& +47$^{\circ}$29$'$40$''$ & 167$\pm$11& 84$\pm$8 & 53.5 & 895 & Sc & 9.91 & 83 & 3600  & 66.08 \\
NGC 4088 & 12h05m34s& +50$^{\circ}$32$'$21$''$ & 173$\pm$14& 62$\pm$7 & 76 & 757 & SABc & 9.51 & 71 & 3600  & 58.13 \\
NGC 3949 & 11h53m41s& +47$^{\circ}$51$'$32$''$ & 164$\pm$7& 63$\pm$5 & 36 & 800 & Sbc & 10.23 & 56 & 4000  & 57.15 \\
NGC 4157 & 12h11m04s& +50$^{\circ}$29$'$05$''$ & 185$\pm$10& 71$\pm$6 & 35 & 774 & SABb & 9.93 & 90 & 4000  & 63.30 \\
NGC 3769 & 11h37m44s& +47$^{\circ}$53$'$35$''$ & 122$\pm$8& 80$\pm$7 & 29 & 737 & Sb & 11.08 & 78 & 8000  & 66.25\\
NGC 3726 & 11h33m21s& +47$^{\circ}$01$'$45$''$ & 162$\pm$9& 63$\pm$10
& 84 & 866 & Sc & 9.50 & 49 & 6600 & 38.44\\ \hline 
\end{tabular}}
\caption{Column 1 gives the galaxy's name, while columns 2 \& 3 give
the J2000 coordinates. The asymptotic circular velocity $v_c$ is
tabulated in column 4, it is taken directly from Verheijen (2001) for
the HSB galaxies, and measured from the published rotation curves for
the LSB galaxies.  Central velocity dispersion ($\sigma_c$) at $r_e/8$
and the effective radius of the galaxy can be found in columns 5 and
6.  The heliocentric radial velocity, Hubble classification, total $I$
band magnitude, inclination can be found in columns 7 to 10 and are
taken from Verheijen (2001) for the HSB galaxies and Palunas \&
Williams (2000) for the LSB galaxies and the online NED
database. Finally, the total exposure time and S/N/pixel at 8500\AA\
are given in columns 11 and 12.\label{list}}
\end{table*}

\section*{Acknowledgments}
We would like to thank S. De Rijcke for sharing his software to derive
velocity dispersion profiles with us, P. Palunas for providing us the
rotational velocities in tabular form and A. Pizzella for the very
fruitful discussions. PB wishes to thank the Herzberg Institute of
Astrophysics for the hospitality which made this investigation
possible. Based on observations made at the European Southern
Observatory, Chile (ESO Programme No. 073.B-0780). Based on
observations collected at the Centro Astron\'omico Hispano Alem\'an
(CAHA) at Calar Alto, operated jointly by the Max-Planck Institut
f\"ur Astronomie and the Instituto de Astrof\'isica de Andaluc\'ia
(CSIC). PB acknowledges the Fund for Scientific Research Flanders
(FWO) for financial support. This research has made use of the
NASA/IPAC Extragalactic Database (NED) which is operated by the Jet
Propulsion Laboratory, California Institute of Technology, under
contract with the National Aeronautics and Space Administration.

\bsp \label{lastpage} 
\begin{thebibliography}{99}
\bibitem[Adams, Graff \& Richstone (2000)]{adams} Adams, F.C., Graff, D.S., Richstone, D., 2000, \apj, 551, L31
\bibitem[Baes et al. (2003)]{baes}  Baes, M., Buyle, P., Hau, G.K.T., Dejonghe, H., 2003, \mnras, 341, L44
\bibitem[Baggett et al. (1998)]{baggett} Baggett, W. E., Baggett, S. M., Anderson, K. S. J., 1998, \aj, 116, 1626
\bibitem[Cappellari \& Emsellem (2004)]{cappellari} Cappellari, M. \& Emsellem, E., 2004, PASP, 116, 138
\bibitem[Cattaneo, Haehnelt \& Rees (1999)]{cattaneo} Cattaneo, A., Haehnelt, M.G., Rees, M.J., 1999, \mnras, 308, 77
\bibitem[Courteau (1996)]{courteau}  Courteau, S., 1996, \apjs, 103, 363
\bibitem[De Bruyne et al. (2003)]{veronique}  De Bruyne, V., Vauterin, P., De Rijcke, S., Dejonghe, H., 2003, \mnras, 339, 215
\bibitem[De Bruyne et al. (2004)]{veronique2}  De Bruyne, V., De Rijcke,
S., Dejonghe, H., Zeilinger, W.W. 2004, \mnras, 349, 461
\bibitem[De Rijcke et al. (2005)]{sven}  De Rijcke, S., Michielsen, D., Buyle, P., Zeilinger, W.W., Dejonghe, H., Hau, G.K.T., 2005, AN, 326, 542
\bibitem[Di Matteo et al. (2003)]{tiziana} Di Matteo, T., Croft, R. A.C., Springel, V., Hernquist, L. 2003, ApJ, 593, 56
\bibitem[Ferrarese (2002)]{laura1}  Ferrarese, L., 2002, \apj, 578, 90
\bibitem[Ferrarese \& Merritt (2000)]{laura2}  Ferrarese, L. \& Merritt, D., 2000, \apjl, 539, L9
\bibitem[Ferrarese \& Ford (2005)]{laura3}  Ferrarese, L., \& Ford, H.C. 2005, Sp.Sc.Reviews, 116, 523
\bibitem[Gebhardt et al. (2000)]{geb}  Gebhardt K. et al., 2000, \apj, 539, L13
\bibitem[Ghez et al. (2003)]{ghez} Ghez, A.M., et al. 2003, ApJ, 586, L127
\bibitem[Graham et al. (2001)]{graham} Graham, A.W., Erwin, P., Caon, N., \& Trujillo, I., 2001, \apj, 563, L11
\bibitem[Haehnelt, Natarajan \& Rees (1998)]{haehnelt1} Haehnelt, M.G.,Natarajan, P., Rees, M.J., 1998, \mnras, 300, 817
\bibitem[Haehnelt \& Kauffmann (2000)]{haehnelt2} Haehnelt, M.G., \& Kauffmann, G., 2000, \mnras, 318, L35
\bibitem[Harmes et al. (1994)]{harms} Harmes, R.J., et al. 1995, \apj,  435, L35 
\bibitem[Heraudeau et al. (1996)]{heraudeau} Heraudeau, P., Simien, F., Mamon, G. A., 1996, \aap, 117, 417
\bibitem[Hopkins et al. (2005)]{hopkins} Hopkins, P. F., Hernquist, L., Martini, P., Cox, T. J., Robertson, B., Di Matteo, T., Springel, V. , 2005, \apjl, 625, L71
\bibitem[Kawakatu et al. (2005)]{kawakatu05} Kawakatu, N., Saitoh, T.R., Wada, K. 2005, \apj, 628, 129 
\bibitem[Kelson (2003)]{kelson} Kelson, D., 2003, PASP, 115, 688
\bibitem[Kormendy \& Richstone (1995)]{kormendy} Kormendy, J. \& Richstone, D., 1995, ARA\&A, 581
\bibitem[Loeb \& Rasio (1994)]{loeb} Loeb, A. \& Rasio, F., 1994, \apj, 432, L52
\bibitem[Marconi \& Hunt (2003)]{marconi} Marconi, A., \& Hunt,  L.K. 2003, ApJ, 589, L21
\bibitem[Miyoshi et al. (1995)]{miyoshi} Miyoshi, M., Moran, J., Herrnstein, J., Greenhill, L.,  Nakai, N., Diamond, P., \& Inoue, M. 1995, Nature, 373, 127
\bibitem[M\"ollenhoff \& Heidt (2001)]{mollenhoff} M\"ollenhoff, C., Heidt, J., 2001, \aap, 368, 16
\bibitem[Monaco, Salucci \& Danese (2000)]{monaco} Monaco, P., Salucci, P., Danese, L., 2000, \mnras, 311, 279
\bibitem[Palunas \& Williams (2000)]{palunas}  Palunas, P. \& Williams, T.B., 2000, \aj, 120, 2884
\bibitem[Pizzella et al. (2005)]{pizzella}  Pizzella, A., Corsini, E.M., Dalla Bont\'a, E., Sarzi, M., Coccato, L., Bertola, F., 2005, ApJ, 631, 785
\bibitem[Press et al. (2002)]{press92} Press, W.H., Teukolsky, S.A., Vetterling, W.T., Flannery, B.P., 2002, Numerical Recipes in C++. Cambridge Univ. Press, Cambridge 
\bibitem[Robertson et al. (2005)]{robertson}  Robertson, B., Hernquist, L., Cox, T.J., Di Matteo, T., Hopkins, P.F., Martini, P., Springel, V., 2005, astro-ph/0506038
\bibitem[Sch\"{o}del et al. (2003)]{schodel} Sch\"{o}del, R., Ott, T., Genzel, R., Eckart, A., Mouawad, N., \& Alexander, T.\ 2003, ApJ, 596, 1015
\bibitem[Shankar et al. (2006)]{shankar06}  Shankar, F., Lap, A., Salucci, P., De Zotti, G., Danese, L. 2006, astro-ph/0601577
\bibitem[Silk \& Rees (1998)]{silk}  Silk, J. \& Rees, M.J.., 1998, \aap, 331, L1
\bibitem[Stuart \& Wyithe (2006)]{stuart06}  Stuart, J. \& Wyithe, B. 2006, \mnras, 365, 1082
\bibitem[Tremaine et al. (2002)]{trem} Tremaine, S. et al., 2002, \apj, 574, 740
\bibitem[Van Zee et al. (2004)]{vzee04} Van Zee, L., Skillman, E.D., Haynes, M.P., 2004, \aj, 128, 121 
\bibitem[Verheijen (2001)]{verheijen1} Verheijen, M., 2001, \apj, 563, 694 
\bibitem[Wyithe \& Loeb (2005)]{wyithe05} Wyithe, J.S. \& Loeb, A. 2005, \apj, 634, 910
\end{thebibliography}
\end{document}